\documentclass[aps,prl,twocolumn,superscriptaddress,showpacs]{revtex4}
\usepackage{graphicx}

\begin{document}


\title{Vortex sorter for Bose-Einstein condensates}


\author{Graeme Whyte}
\affiliation{Department of Physics and Astronomy, University of Glasgow, Glasgow G12~8QQ, United~Kingdom}

\author{John Veitch}
\affiliation{Department of Physics and Astronomy, University of Glasgow, Glasgow G12~8QQ, United~Kingdom}

\author{Patrik \"Ohberg}
\affiliation{Department of Physics, University of Strathclyde, Glasgow G4~0NG, United~Kingdom}

\author{Johannes Courtial}
\affiliation{Department of Physics and Astronomy, University of Glasgow, Glasgow G12~8QQ, United~Kingdom}
\email[]{j.courtial@physics.gla.ac.uk}


\date{\today}

\begin{abstract}
We have designed interferometers that sort Bose-Einstein condensates into their vortex components.  The Bose-Einstein condensates in the two arms of the interferometer are rotated with respect to each other through fixed angles; different vortex components then exit the interferometer in different directions.  The method we use to rotate the Bose-Einstein condensates involves asymmetric phase imprinting and is itself new.  We have modelled rotation through fixed angles and sorting into vortex components with even and odd values of the topological charge of 2-dimensional Bose-Einstein condensates in a number of states (pure or superposition vortex states for different values of the scattering length).  Our scheme may have applications for quantum information processing.
\end{abstract}




\maketitle


\noindent
\textit{Introduction.} One of the central characteristics of a superfluid such as a Bose-Einstein condensate (BEC) is the presence of quantized vortices.
Vortices have been  generated experimentally \cite{Madison-et-al-2000, Abo-Shaeer-et-al-2001} by stirring the BEC, very similar to the rotating-bucket experiment in Helium~\cite{Tilley-Tilley-1990}.  The detection of vortices in BECs is typically made by a direct observation of the vortex core or by interference experiments~\cite{Chevy-et-al-2001}.  In present experiments the charge, $m$, of the vortex is in principle known as the initial rotation frequency which stirs the cloud is known.  In this paper we show how to sort vortices when the charge is not known.

Vortices have attracted considerable interest both experimentally and theoretically, mainly because of their inherent many-body character and the connection to fluid dynamics.  In addition, optical vortices in single photons have recently been used to carry information -- and in particular quantum information \cite{Molina-Terriza-et-al-2002}.
Light is an excellent carrier of information over large distances as the photons travel very fast and do not easily interact with each other.  For the same reason, photons are not very well suited for storing the information for longer times.  This is where atoms would be better suited as a medium for storing information, especially quantum information.
Optical vortices, special cases of light with orbital angular momentum, can carry huge amounts of information as there is in principle no limit to the quantized angular momentum acting as the information carrier. Transferring this angular momentum to atoms would constitute a way to store the information~\cite{Lukin-2003}. If such a storage device is to work with atoms we need a way to manipulate atomic states, and in particular vortex states, in an efficient and useful way.  It is therefore important to know the mechanisms behind the vortex dynamics and more importantly how to manipulate the vortex states in order to be able to make any kind of readouts from the trapped quantum gas.  In this paper we study theoretically the application to BECs of ideas borrowed from optical vortex sorting~\cite{Leach-et-al-2002}.


\noindent
\textit{Vortex sorter.}  If a vortex (in light or in a BEC) of charge $m = 1$ is rotated through $180^\circ$ about its centre, it changes phase by $\pi$ (and, in the simplest case, is unchanged in any other respect).  If, on the other hand, a vortex of charge $m = 2$ is rotated through $180^\circ$, its phase is unchanged.  The two cases discussed above are in fact representative for all vortices with odd and even charges, respectively.  This effect has been used in an optical two-arm interferometer which rotates the beams in the two arms with respect to each other to route vortices according to their charge into one of the interferometer's two exit ports \cite{Leach-et-al-2002}.    When the beams are re-combined, even-charge vortex components interfere constructively in one interferometer port and therefore exit the interferometer through it, while odd-charge vortex components interfere destructively in that port and therefore exit the interferometer through another port (in which even-charge vortices interfere destructively).  The vortices exiting from the two ports can be sorted further in similar interferometers, but with different relative rotation angles \cite{Leach-et-al-2002}.  For example, vortices with even charges ($m = 0, \pm 2, \pm 4, ...$) can be sorted into those whose charges are respectively integer and half-integer multiples of 4 ($m \text{ mod } 4 = 0$ or $m \text{ mod } 4 = 2$, respectively).  In some cases, uniform phase offsets in one arm are required~\cite{Wei-et-al-2003}.

\begin{figure}
\includegraphics[width=8cm]{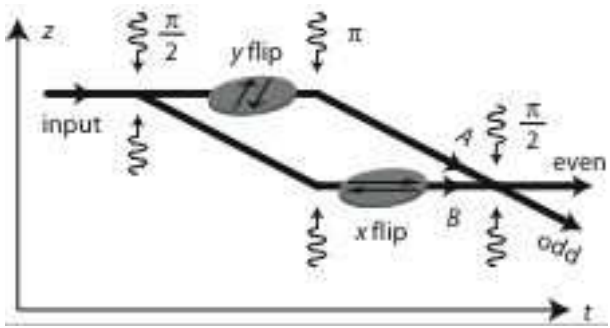}
\caption{Schematic of a vortex-sorting Bragg-pulse interferometer.  A stationary input BEC is first split by a $\pi/2$ Bragg pulse (left; the horizontal axis is time).  The two resulting BECs are in different momentum states and move apart.  A $\pi$ Bragg pulse swaps the momentum states, so that the two BECs move together again.  We refer to the two different trajectories in this space-time diagram as two `arms' of the interferometer.  The BECs in the two arms are flipped vertically and horizontally, respectively, which corresponds to a relative rotation through $180^\circ$.  A second $\pi/2$ Bragg pulse mixes the two BECs such that even and odd vortex components have different momentum states -- they exit the interferometer through different `ports'.}
\label{interferometer-figure}
\end{figure}

By using Bragg pulses, it is possible to coherently split and re-combine a BEC as would be required within a two-arm interferometer (figure \ref{interferometer-figure})\cite{Giltner-et-al-1995}, and using specially designed light pulses a BEC could be rotated through any given angle (see below).  We numerically examine here a vortex sorter created by combing these two elements as shown in figure \ref{interferometer-figure}.

Specifically, we model a 2-dimensional BEC that is split into two identical BECs, which are then rotated with respect to each other through $180^\circ$, and finally superposed. The wave function of the original BEC, $\Psi$, is split according to
\begin{equation}
\Psi_1 = \Psi_2 = \frac{1}{\sqrt{2}} \Psi.
\label{BEC-splitting}
\end{equation}
The two BECs are rotated with respect to each other through mirroring the two BECs with respect to the $x$ and $y$ axis, respectively.
As explained below, this can be achieved through imprinting specific phases onto the BECs at specific times.  In combination with the time evolution between the phase-imprinting events, this results in wave functions $\Psi^\prime_1$ and $\Psi^\prime_2$.
We simulate this time evolution according to the time-dependent Gross-Pitaevskii equation \cite{Pethick-Smith-2002}, which we write in the form \cite{Caradoc-Davies-2000}
\begin{equation}
i \hbar \frac{ \partial \psi }{\partial t} =
\left[ -\frac{\hbar^{2}}{2m} \nabla^2 + \frac{1}{2} m \omega^2 r^2 + g |\psi|^2 \right] \psi,
\end{equation}
where $m$ is the mass of each atom, $\omega$ is the trap frequency and $g$ is the non-linear coefficient, which is given by
\begin{equation}
g = 4 \pi N \frac{a}{d} \frac{\hbar^2}{m}.
\end{equation}
Here $d$ is the effective thickness of the BEC in $z$ direction and $N$ is the number of atoms. The motion can be considered two-dimensional if the chemical potential $\mu$ of the trapped cloud is smaller than the corresponding trapping energy $\hbar\omega_z$ in the $z$ direction.
Finally, the two wave functions $\Psi^\prime_1$ and $\Psi^\prime_2$ are superposed according to the equations
\begin{equation}
\Psi_{\text{even}} = \frac{1}{\sqrt{2}} (\Psi^\prime_1 + \Psi^\prime_2), \quad
\Psi_{\text{odd}} = \frac{1}{\sqrt{2}} (\Psi^\prime_1 - \Psi^\prime_2).
\label{BEC-recombination}
\end{equation}

This model can represent various interferometers, all of which are idealised in some respects.  For example, a Bragg-pulse interferometer with rotation in the arms (figure \ref{interferometer-figure}), is idealised as follows.  Firstly, the Bragg pulses are assumed to be perfect, that is acting according to equations (\ref{BEC-splitting}) and (\ref{BEC-recombination}), which describe perfect $\pi/2$ pulses with the exception that the two states $\Psi_1$ and $\Psi_2$ have different momenta.  The $\pi$ pulses, which are also required in the Bragg-pulse-interferometer scheme, and which swap the BECs between the two states, are also assumed to be perfect (which is consistent with experiments in which fringe visibilities close to 1 were achieved in Bragg-pulse interferometers \cite{Torii-et-al-2000}).  Secondly, the interaction between the BECs in the different arms is neglected.  To the best of our knowledge, no experimentally realisable situation is exactly represented by this, but some are represented better than others.  A Bragg-pulse direction that separates the planes of the two BECs, for example, should lead to less interaction between the BECs than a Bragg-pulse direction that move the two planes across each other; however, applying the light pulses for rotation to the two arms separately is potentially difficult in this geometry.  Thirdly, the arm length is just that required for rotation; we have made no allowance for any additional time it might take for the BECs to separate sufficiently such that they can be rotated independently and subsequently recombined.  However, in analogy to optical imaging, lens light pulses \cite{Whyte-et-al-2003} might be able to return the BEC into an earlier state, thereby effectively shortening the arms.  Other interferometer types that approximate our idealised model include, for example, those that split a BEC into two by putting them into different internal states \cite{Chevy-et-al-2001} and manipulate the two BECs independently through phase imprinting with light pulses with different detunings.

\begin{figure}
\includegraphics[width=8cm]{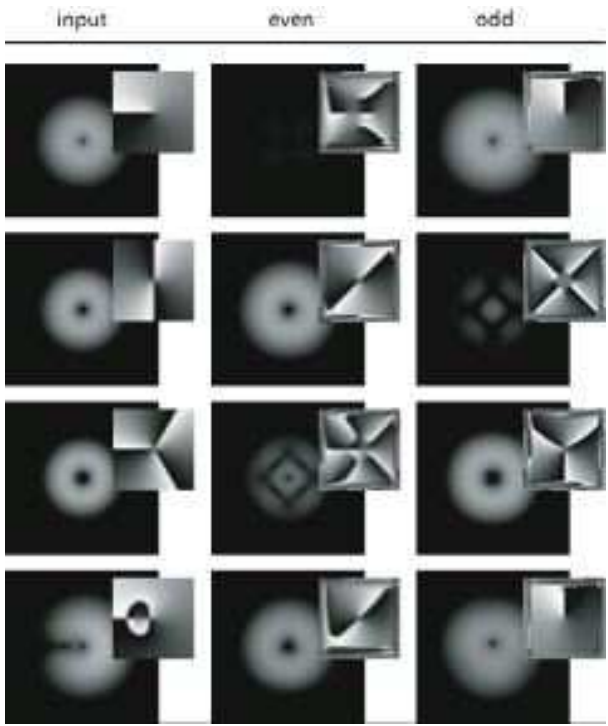}
\caption{Simulated vortex sorting in an idealised interferometer.  The columns show the probability density and phase (inset at reduced size) of the input BEC (left) and the BEC in the `even' (centre) and `odd' (right) output ports.  The top three cases show sorting of pure vortices in $m = 1$, $m = 2$, and $m = 3$ states; depending on the value of $m$ as it enters the interferometer, the BEC comes either out of the even or odd output port.  This sorting is not perfect:  a small fraction can be seen to come out of the `wrong' output port.  The fourth case shows sorting of a superposition of vortices with charges $m = 1$ and $m = 2$, which get split into its vortex components.  This figure was calculated for $g = 0$.  Note that most of the structure near the edge of most phase plots in this paper, where the probability density is very low, is a numerical artefact.}
\label{interfered-BEC-figure}
\end{figure}

\begin{figure}
\includegraphics[width=8cm]{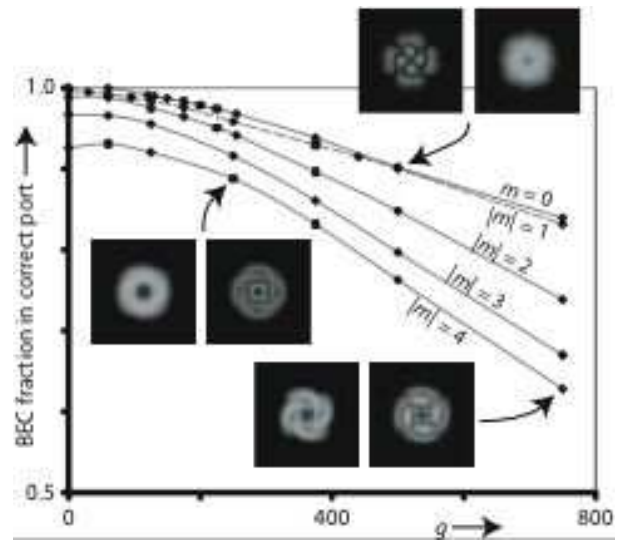}
\caption{Fraction of the BEC in the correct port as a function of the non-linear coefficient, $g$, and for different values of the topological charge.  Inset are the density cross-sections in the even (left) and odd (right) ports corresponding to some of the data points ($m = 1$, $g = 500$ and $m = 4$, $g = 250$ and $750$).}
\label{correct-fraction-figure}
\end{figure}

Figure \ref{interfered-BEC-figure} shows examples of sorting a BEC into its `even' and `odd' vortex components, which is demonstrated for pure vortex states as well as superpositions.   These simulations -- indeed all the simulations in this paper -- were performed over an area of $14 \times 14$ (in units of $\sqrt{\hbar/m\omega}$) on a $256 \times 256$ grid of wave function amplitudes.  Figure \ref{correct-fraction-figure} shows the fraction of the original BEC that exits the interferometer in the correct port -- a measure of the quality of the sorting -- as a function of the non-linear coefficient $g$.  It can be seen that the scheme works better for small values of $g$.  

\noindent
\textit{Rotation of BECs.} Several methods already exist for rotating a BEC through a given angle, using, for example, an external magnetic field \cite{Marzlin-Zhang-1998b}, a careful arrangement of laser beams \cite{Marzlin-Zhang-1998}, or Bragg pulses \cite{Giltner-et-al-1995,Poulsen-Molmer-2002}.  We describe here a novel method based on an optical analogy \cite{Beijersbergen-et-al-1993}.

\begin{figure}
\includegraphics[width=8cm]{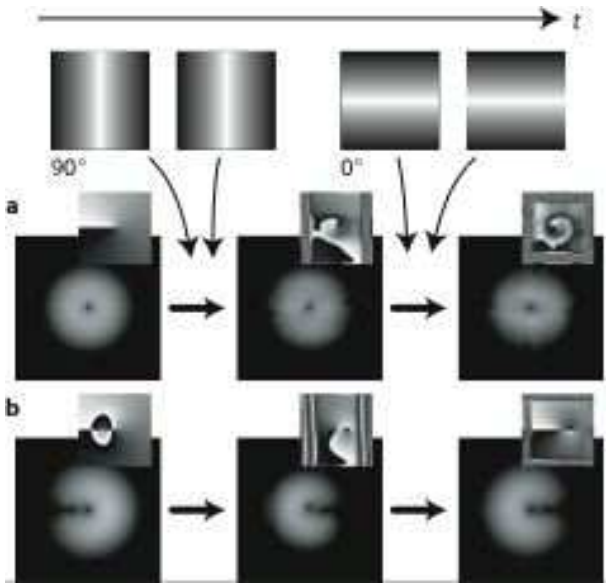}
\caption{Rotating a BEC through $180^\circ$ with a series of cylindrical-lens pulses.  Shown at the top are the intensity cross sections of four far off-resonant light pulses.  The intensity of the first two pulses falls of quadratically in the $y$, that of the other two pulses falls of quadratically the $x$ direction.  The pictures below show examples of simulated BECs before interaction with the light pulses (left), after interaction with the first pair of pulses (centre), and after interaction with the second pair (right).  \textbf{a}: pure vortex with $m = 1$, $g = 500$.  It can be seen that the handedness of the phase distribution (inset) is reversed after the first pair of lens pulses, and that it is back to its original handedness after the second pair.  That this leads to rotation can be seen more clearly in \textbf{b}, which starts off with a superposition of $m = 1$ and $m = 2$ vortices ($g = 0$): after interaction with all pulses the probability density is indeed rotated through $180^\circ$.
In both cases, the focal time of each cylindrical-lens pulse was $t_f = 0.03$ (in units of $1/\omega$); the pulses passed the BECs at times $t = 0$, $0.06$, $0.06$ (the two middle pulses were merged into one), and $0.12$, respectively.}
\label{rotation-figure}
\end{figure}

Our method is based on the fact that mirroring at one axis (or plane in 3 dimensions), followed by mirroring at another axis, which is rotated with respect to the first axis by an angle $\alpha$, is equivalent to a rotation through an angle $2 \alpha$ about the intersection between the two axes.  In analogy to mirroring of a light beam, which can be achieved with a pair of identical cylindrical lenses parallel to the mirror axis, each of focal length $f$, which are separated by $2 f$ (such a configuration is called a $\pi$ mode converter \cite{Beijersbergen-et-al-1993}), a BEC can be mirrored by a pair of correctly spaced cylindrical-lens pulses.  These are far off-resonant light pulses with a transverse intensity distribution that is proportional to the thickness of the corresponding optical cylindrical lenses, that is the intensity falls off quadratically in one direction and is constant in the other.  The effect of each cylindrical-lens pulse is a phase change proportional to the local intensity \cite{Dobrek-et-al-1999,Denschlag-et-al-2000}:  the cylindrical-lens pulses act like phase holograms of cylindrical lenses~\cite{Whyte-et-al-2003}.  The phase change due to each lens pulse is $r^2 / (4 t_f)$, where $r$ is the distance from the axis of the cylindrical-lens pulse and $t_f$ is its focal time (the equivalent of the focal length in optical lenses).  In this paper we use $t_f = 0.03$ (in units of $1/\omega$), which is one of the smallest focal times that satisfies the Nyquist criterion for our model, and a time of $t_d = 0.06$ between the lens pulses.  Figure \ref{rotation-figure} illustrates modelled examples of rotation of BECs through $180^\circ$.

This scheme does not work perfectly, not even in optics: a light beam (and, by analogy, a BEC with $g = 0$) is mirrored perfectly only in the limit of cylindrical lenses with infinitely short focal lengths \cite{Beijersbergen-et-al-1993}.  Obviously, this is not possible, and the result is imperfect mirroring that leads to asymmetry and vortex splitting.  Another problem when using cylindrical-lens pulses to mirror BECs with $g \neq 0$ is that the BEC can intermittently become focussed into a line, which greatly amplifies the non-linear effects, which in turn usually lowers the quality of the mirroring.

In the context of the vortex-sorting interferometer, we are interested in differential rotation between the BECs in the two arms of the Bragg-pulse interferometer.  A better way of achieving such differential rotation is to apply the first two of the four rotation pulses shown in figure \ref{rotation-figure} to the BEC in one arm, and the other two to the BEC in the other arm; both BECs are mirrored, but with respect to different axes.  As demonstrated above, we find that -- in the spirit of spreading imperfections symmetrically and hoping for cancellation -- this leads to good results for $g \gtrsim 50$ for small values of $|m|$.


\noindent
\textit{Conclusions.} In this paper we have investigated the sorting of vortices in BECs using an interferometric technique. Our technique requires the BEC to be rotated, which we achieve with spatially inhomogeneous imprinted phases.  If the non-linearity is strong, the efficiency of the rotation and therefore the efficiency of the sorting process are decreased, but with existing experimental techniques such as exciting Feshbach resonances it should be possible to ``tune'' the non-linear coefficient $g$ to a value suitable for reliable vortex sorting.

The techniques presented here are based on methods used in conventional optics.  When these methods are transferred to BECs, complications arise, but also some intriguing new possibilities.  In particular non-linearity -- the origin of most complications -- is important whenever information is not only to be stored, but also to be processed in computations.

\begin{acknowledgments}
Thanks to Aidan Arnold for very useful discussions.  JC is a Royal Society University Research Fellow, P\"O is a Royal Society of Edinburgh Research Fellow.  Financial support is also gratefully acknowledged from the UK Engineering and Physical Sciences Research Council (EPSRC).
\end{acknowledgments}


\end{document}